\global\def\draftcontrol{0} 

\def\versionno{ time -- draft -- 11.13.03  }

\catcode`\@=11

\expandafter\ifx\csname draftcontrol\endcsname\relax\global\def\draftcontrol{0}
\fi

{\count255=\time\divide\count255 by 60
\xdef\hourmin{\number\count255}
\multiply\count255 by-60\advance\count255 by\time
\xdef\hourmin{\hourmin:\ifnum\count255<10 0\fi\the\count255}}
\def\draftdate{\number\month/\number\day/\number\year\ \ \ \hourmin }

\newcommand\makepapertitle{\par
  \begingroup
    \renewcommand\thefootnote{\@fnsymbol\c@footnote}%
    \def\@makefnmark{\rlap{\@textsuperscript{\normalfont\@thefnmark}}}%
    \long\def\@makefntext##1{\parindent 1em\noindent
            \hb@xt@1.8em{%
                \hss\@textsuperscript{\normalfont\@thefnmark}}##1}%
     \newpage
     \global\@topnum\z@   
     \@makepapertitle
     \thispagestyle{empty}\@thanks
  \endgroup
  \setcounter{footnote}{0}%
  \global\let\thanks\relax
  \global\let\makepapertitle\relax
  \global\let\@makepapertitle\relax
  \global\let\@thanks\@empty
  \global\let\@author\@empty
  \global\let\@date\@empty
  \global\let\@title\@empty
  \global\let\title\relax
  \global\let\author\relax
  \global\let\date\relax
  \global\let\and\relax
  \def\version{\let\version\@version\@gobble}
}
\def\@makepapertitle{%
  \newpage
   \ifnum\draftcontrol=1 {}
   \version\versionno
   \vskip 3em%
   \else
   \hfill\hbox to 3cm {\parbox{4cm}{\@pubnum}\hss}%
   \vskip 3em%
   \fi
   \begin{center}%
   \let \footnote \thanks
     {\LARGE \@title \par}%
     \vskip 1.5em%
     {\normalsize
       \lineskip .5em%
       \begin{tabular}[t]{c}%
         \@author
       \end{tabular}\par}%
     \vskip 1em%
     {\@bstract}%
     \end{center}%
     \vskip .5em
     \@date%
   \par
}

\gdef\@pubnum{}
\def\pubnum#1{%
  \gdef\@pubnum{#1}}

\gdef\@bstract{}
\def\Abstract#1{%
  \gdef\@bstract{%
   \parbox{\textwidth-0pc}{%
   \centerline{\bf Abstract}\penalty1000
   \noindent
   \renewcommand\baselinestretch{1.0}
   {#1}}}
}

\def\ps@paper{\let\@mkboth\@gobbletwo%
     \ifnum\draftcontrol=1
        \def\@oddfoot{\hbox to \textwidth{\tiny \versionno \hfil\tiny\draftdate}%
        \hskip -\textwidth \hbox to \textwidth{\hfil\rm\thepage\hfil}}%
     \else\def\@oddfoot{\hbox to \textwidth{\hfil\rm\thepage\hfil}}
     \fi
     \let\@evenfoot\@oddfoot
}

\def\body{\clearpage
          \pagestyle{paper}
        }
\newenvironment{acknowledgments}{%
\vskip 3.25ex
\noindent {\bf Acknowledgments}
}

\def\@version#1{\ifnum\draftcontrol=1
\typeout{}\typeout{#1}\typeout{}
\vskip3mm\centerline{\hbox{\fbox{\normalsize{\tt DRAFT -- #1 -- }
                   {\draftdate}}}}\vskip3mm
\fi}
\let\version\@version
\long\def\eqlabel#1{\ifnum\draftcontrol=1
                    \tag@false  
                    \tag*{(\theequation) \hbox to -0.2cm{\hspace{0cm}\small{#1}\hss}}
                    \refstepcounter{equation} 
                    \edef\@currentlabel{\theequation}
                    \ltx@label{#1}          
                    \else
                    \label{#1}
                    \fi
                    }

\documentclass[12pt,letterpaper]{article}

\usepackage{amsmath,amssymb,array,calc,epsfig}

\ifnum\draftcontrol=1
\fi

\renewcommand\baselinestretch{1.25}
\renewcommand\section{\@startsection {section}{1}{\z@}%
                                   {-3.5ex \@plus -1ex \@minus -.2ex}%
                                   {2.3ex \@plus.2ex}%
                                   {\normalfont\large\bfseries}}
\renewcommand\subsection{\@startsection{subsection}{2}{\z@}%
                                     {-3.25ex\@plus -1ex \@minus -.2ex}%
                                     {1.5ex \@plus .2ex}%
                                     {\normalfont\normalsize\bfseries}}
\renewcommand\subsubsection{\@startsection{subsubsection}{3}{\z@}%
                                     {-3.25ex\@plus -1ex \@minus -.2ex}%
                                     {1.5ex \@plus .2ex}%
                                     {\normalfont\normalsize\it}}

\numberwithin{equation}{section}

\def\complex      {{\mathbb C}}

\def\projective   {{\mathbb P}}

\def\reals        {{\mathbb R}}
\def\zet          {{\mathbb Z}}

\def\be     {\begin{equation}}
\def\ende       {\end{equation}}

\def\revise#1       {\marginpar{\rule{2mm}{1cm} #1}}

\def\ZZ{\zet}

\def\PP{\projective}

\newcommand{\nc}{\newcommand}

\def\bea        {\begin{eqnarray}}
\def\eea        {\end{eqnarray}}
\nc{\e}{{\rm exp}}

\nc{\cosech}{{\rm cosech}}

\nc{\Li}{{\rm Li_{2}}}

\nc{\li}{\lambda_{i}}
\nc{\lj}{\lambda_{j}}
\nc{\lk}{\lambda_{k}}
\nc{\laml}{\lambda_{l}}

\nc{\mi}{\mu_{i}}
\nc{\mj}{\mu_{j}}
\nc{\mk}{\mu_{k}}
\nc{\ml}{\mu_{l}}

\nc{\om}{\omega}

\def\O{{\cal O}}

\nc{\ra}{\rightarrow}

\nc{\defconi}{T^{*}(S^{3})}
\nc{\Sthree}{S^{3}}
\nc{\Sthreep}{S^{3}/\ZZ_{p}}
\nc{\defconip}{T^{*}(S^{3}/\ZZ_{p})}
\nc{\resconi}{\O_{-1} + \O_{-1} \ra \PP^{1}}

\nc{\non}{\nonumber}

\catcode`\@=12

\begin{document}

\title{Large $N$ Duality, Lens Spaces and the Chern-Simons Matrix Model} 

\pubnum{
CALT-68-2468 \\
USC-03-08 \\
NSF-KITP-03-110 \\
hep-th/0312145}

\date{December 2003}

\author{Nick Halmagyi\footnote{halmagyi@physics.usc.edu}$\ ^{1,2}$, 
Takuya Okuda\footnote{takuya@theory.caltech.edu}$\ ^{3}$
and Vadim Yasnov\footnote{yasnov@physics.usc.edu}$\ ^{1}$\\[0.4cm]
\it $^{1}$Department of Physics and Astronomy\\
\it University of Southern California \\
\it Los Angeles, CA 90089, USA \\[0.2cm]
\it $^{2}$Kavli Institute for Theoretical Physics\\
\it University of California\\
\it Santa Barbara, CA 93106, USA \\ [0.2cm]
\it $^{3}$California Institute of Technology \\
\it Pasadena, CA, 91125, USA\\
}

\Abstract{
We demonsrate that the spectral curve of the matrix model for Chern-Simons theory on the Lens space $S^{3}/\ZZ_p$ is precisely the Riemann surface which appears in the mirror to the blownup, orbifolded conifold. This provides the first check of the $A$-model large $N$ duality for $T^{*}(S^{3}/\ZZ_p)$, $p>2$.
}

\enlargethispage{1.5cm}

\makepapertitle

\body


\section{Introduction}

The conifold transition is an example of an open/closed string duality. In the topological $A$-model, this is a duality between the open $A$-model on $\defconi$ (which is equivalent to Chern-Simons (CS) theory on $\Sthree$ \cite{Witten:1992fb}), and Kahler gravity on $\resconi$. This was originally studied by taking the partition function of large $N$ Chern-Simons (CS) theory on $\Sthree$ expanded in a 't Hooft limit and presenting it first in the form of an open string theory \cite{Gopakumar:1998ii}(i.e. an expansion in genus and holes) and then summing over the holes to get a closed string theory \cite{Gopakumar:1998ki}. A worldsheet proof of this duality has since been provided \cite{Ooguri:2002gx}. 

It is of some interest to extend this to a duality between $\defconip$ and a $\ZZ_{p}$ orbifold of the resolved conifold. Whilst an explicit form of the partition function for CS theory on $\Sthreep$ is known \cite{Witten:1988hf}, this form includes the summation over all vacua in the theory, yet for the purposes of string theory we want only the contribution from a single vacuum. For perhaps these reasons it has not been possible to exhibit the large $N$ duality of $T^{*}(\Sthreep)$ at the level of partition function however the worldsheet proof does generalize to $\defconip$ and other geometries \cite{Takuya}.

In \cite{Marino:2002fk, Aganagic:2002wv} it was shown that CS theory on $\Sthreep$ has a matrix model description. In \cite{Aganagic:2002wv} it was also shown that Holomorphic Chern-Simons (HCS) theory reduced to $\PP^{1}$'s inside the mirror (call it $\widetilde{X}$) to $\defconip$ has a matrix model description. Further, these matrix models are identical. For the case $p=2$ the partition function of this matrix model was calculated perturbatively and was shown to agree with the Kodaira-Spencer theory \cite{Bershadsky:1993cx} predictions from the large $N$ dual geometry, providing solid evidence for the proposed duality. For CS theory on $\Sthree$ the matrix model was solved to all genus using orthogonal polynomials in \cite{Tierz:2002jj} and the orientifold of the conifold was studied in \cite{Halmagyi:2003fy}.

The manifold $\widetilde{X}$ is given by the blowup of

\be \label{fp}
xy=F_{p}(e^{u},e^{v}),
\ende
where

\be
F_{p}(e^{u},e^{v})=(e^{v}-1)(e^{v+pu}-1)
\ende
and by the general arguments of Dijkgraaf-Vafa theory \cite{Dijkgraaf:2002fc, Dijkgraaf:2002dh}, the spectral curve of the corresponding matrix model should be a complex structure deformation of $F_p=0$. In \cite{Halmagyi:2003ze} two of the current authors found an expression for the spectral curve of the matrix model for CS theory on $\Sthreep$. This involved first showing that although this matrix model looks similar to a $p$-matrix model, it does in fact have square root branch cuts, therefore its spectral curve has only two sheets. This led to an explicit expression for the resolvent, depending on $p-1$ parameters $d_i$ which in principle could be found perturbatively by performing the $A$-cycle integrals. The spectral curve can be read off from the resolvent and the $d_i$ correspond to complex structure moduli. This will be reviewed in section 2.

In section 3 we use toric geometry to construct a resolution of the $\ZZ_p$ orbifold of the resolved conifold. This is a particular $A_{p}$ fibration over $\PP^{1}$. Then using the Hori-Vafa mirror map we can write down the mirror geometry and find that after a suitable co-ordinate redefinition, the non-trivial Riemann surface inside this threefold is precisely the spectral curve found in \cite{Halmagyi:2003ze}. This explicitly identifies the large $N$ dual of $\defconip$ for all $p>1$. 
The matching of the geometries proves the equivalence of the leading order (in $g_s$) free energy between the matrix model and the closed $A$-model on this particular fibration. To our knowledge this is the first check of this large $N$ duality for $p>2$.


\section{The matrix model spectral curve}

The partition function of CS theory on the Lens space $S^{3}/\ZZ_{p}$ \cite{Aganagic:2002wv} can be written as a matrix integral over $p$ sets of eigenvalues, which we label by an index \mbox{$I \in \{0,..,p-1\}$}. The $I$th set contains $N_I$ eigenvalues. The measure is a product of two factors, a self interaction term $(\Delta_1)$ and a term containing the interaction between different sets of eigenvalues $( \Delta_2)$,

\bea
&&\Delta_1 (u)=\prod_I\prod_{i\ne j}{\left (2\sinh\left (\frac{u^I_i-u^I_j}{2}\right )\right )}^2 \\
&&\Delta_2 (u)=\prod_{I<J}\prod_{i,j}{\left (2\sinh\left (\frac{u_i^I-u_j^J+d^{IJ}}{2}\right )\right )}^2,
\eea
where $d^{IJ}=2\pi i (I-J)/p$. The potential has an overall factor of $p$ compared to the $S^{3}$ case,

\be
V(u)=p\sum_{I,i}\frac{{(u_i^I)}^2}{2}.
\ende
In the above notations the CS partition function becomes
\be
Z\sim \int\prod_{I=0}^{p-1}\prod_{i=1}^{N_I}du_i^I\Delta_1 (u)\Delta_2 (u){\rm exp} \left (-\frac{1}{g_s}V(u)\right ).
\ende
We define individual resolvents for each set of the eigenvalues by

\be
\label{resInd}
\omega_I (z)=g_s\sum_i\coth\left (\frac{z-u_i^I}{2}\right )
\ende
and the total resolvent, which we are most interested in is 

\be
\label{resTot}
\omega (z)=\sum_I\omega_I\left (z-\frac{2\pi i I}{p}\right ).
\ende

Anticipating taking the large $N$ limit we also introduce 't Hooft parameters $S_I=g_s N_I$ and $S=\sum_I S_I$.  
The equation of motion for each eigenvalue is 

\be \label{zpeom}
p u_{i}^{I}= g_{s}\sum_{j\neq i}\coth\left( \frac{u^{I}_{i}-u^{I}_{j} }{2} \right)
+ g_{s} \sum_{J\neq I} \sum_j \coth\left( \frac{u^{I}_{i}-u^{J}_{j}+d^{IJ} }{2} \right).
\ende
From the large $N$ limit of this equation we can derive

\be
\label{EOM}
\frac{1}{2}\omega^2 (z)-p\sum_I\left (z-\frac{2\pi i I}{p}\right )\omega_I \left (z-\frac{2\pi i I}{p}\right )=f(z),
\ende
where $f(z)$ is a regular function
\be
f(z)=p~g_s\sum_I\sum_i (u_i^I+2\pi i I/p-z )\coth\left (\frac{z-u_i^I-2\pi i I/p}{2}\right )+\frac{S^2}{2}.
\ende
Given the large $N$ limit of the equations of motion it is possible to find the total resolvent $\omega (z)$. This method has been developed and checked in \cite{Halmagyi:2003ze} which we will now review. We assume that the eigenvalues spread only along the real line. For general multi matrix models this is not true \cite{Dijkgraaf:2002vw, Hofman:2002bi, Lazaroiu:2003vh}. However, this assumption leads to the correct result for our case. Note that we do not make any assumption on the type of the cuts. In the total resolvent $\omega (z)$, the individual resolvents come with relative shifts of the argument by $2\pi i I/p$. Therefore the  cuts in the total resolvent are now separated by $2\pi i I/p$. For example, if $\omega_0 (z)$ jumps at the point $z$, all other individual resolvents $\omega_I (z-2\pi i I/p)$ with $I\ne 0$ are regular at this point. This means that on $I$th cut the total resolvent jumps only due to the resolvent $\omega_I(z)$. From this it follows that

\be
\frac{1}{2}\left( \omega_{+}\left (z+\frac{2\pi i I}{p}\right )+\omega_{-}\left (z+\frac{2\pi i I}{p}\right )\right) =pz,\ \ (I{\rm 'th\ cut})
\ende 
and so every cut is  a square root.

We label the contour around the $I$th cut as $A_I$. From (\ref{resTot}) it is clear that

\bea
&& \lim_{z\rightarrow \infty}\omega(z)=S, \\
\label{2ndsheet}
&& \lim_{z\rightarrow -\infty}\tilde{\omega} (z)=-S,
\eea
where $\tilde{\omega}(z)$ is the value of the resolvent on the second sheet. From (\ref{resInd}) we also have that

\be
\label{Aperiod}
\frac{1}{2}\oint_{A_I}\omega (z)dz=2\pi iS_I.
\ende
Since the integral over the $A=\sum_I A_I$ cycle is fixed by (\ref{2ndsheet}), there are only $p-1$ independent periods.
Now we construct a regular function,

\be\label{g_p}
g(Z)=e^{\omega/2}+Z^pe^{-\omega/2},
\ende
which has the limiting behavior,

\bea
&& \lim_{Z\rightarrow \infty}g(Z)=e^{-S/2}Z^{p}\\
&&  \lim_{Z\rightarrow 0}g(Z)=e^{-S/2}\label{minusInfty}
\eea
and is thus of the form,
\be
g(Z)=e^{-S/2}(Z^p+d_{p-1}Z^{p-1}+...+d_{1}Z+1).
\ende
The function $g(Z)$ depends on $p\! -\!1$ moduli $d_{n}$, which could be found by evaluating the period integrals
(\ref{Aperiod}).  
Since we have already fixed the integral over the cycle $A=\sum_I A_I$ by (\ref{minusInfty}), there are only $p\! -\! 1$ independent $A$-periods. 

We can solve (\ref{g_p}) for $\omega(Z)$ to get

\be \label{zp_w}
\frac{\omega (Z)}{2}=\log\left (\frac{1}{2}\left (g(Z)-\sqrt{g^2 (Z)-4 Z^p}\right )\right),
\ende
the function under the square root is a polynomial of the degree $2p$, it has $2p$ distinct roots that depend on only $p\! -\! 1$ parameters. Thus the spectral curve consists of two cylinders glued together along $p$ cuts. Note that the center of the $I$'th cut is at the point $z=2\pi i I/p$. The cycles around each cut are $A$ cycles. There is also a set of noncompact cycles, the $B$ cycles. Each $B_I$ cycle starts at infinity on the physical sheet (the sheet
where the resolvent is finite) and goes through the $I$'th cut to a point at infinity on the other sheet.

From (\ref{zp_w}) we see that the spectral curve is given by

\be \label{spectral}
(e^{v}-1)(e^{pu+v}-1)+e^{S}-1+e^{v}\sum_{n=1}^{p-1}d_{n}e^{nu}=0,
\ende
which is a complex structure deformation of  $F_{p}=0$ (from (\ref{fp})). Here $z\equiv u$ and $v=(S-\omega )/2$. 
It is worth mentioning  that the functions $d_n(S_I)$ are available as a power series in 't Hooft parameters $S_I$'s which is valid only in the region of small $S_I$. We will see that this region corresponds to negatively large
values of Kahler parameters in the $A$-model. 


\section{The orbifold of the resolved conifold}

We will now construct the $\ZZ_p$ orbifold of the resolved conifold ($\O_{-1} + \O_{-1}\! \ra \! \! \PP^{1}$) using standard toric methods\footnote{a good reference for the basics of toric geometry is \cite{Greene:1996cy}}. We will then apply the Hori-Vafa mirror map, from which we can obtain the Riemann surface which is the non-trivial part of the mirror geometry.

The deformed conifold ($\defconi$) can be written as $\det A=\mu$, where 

\be
A=\left(
\begin{array}{cc}
  z_1 & z_3 \\
  z_2 & z_4 \\
\end{array}
\right).
\ende
We can orbifold this geometry by a $\ZZ_{p}$ symmetry generated by

\be \label{zetpaction}
A\mapsto 
\left(\begin{array}{ll}
\alpha & 0 \\
0 &\alpha^{-1} \\
\end{array}\right)A,
\ende
where $\alpha=e^{2\pi i/p}$. The $\Sthree$ is given by $|z_1|^2+|z_3|^2=\mu$ (assuming $\mu$ is real and positive) and the $\zet_p$ acts on this as $(z_1,z_3)\rightarrow (\alpha z_1, \alpha z_3)$, which gives the Lens space $S^3/\zet_p=L(p,1)$, thus the entire threefold is now $\defconip$.

We now perform this orbifold action on the other side of the proposed large $N$ duality. The resolved conifold can be expressed as

\be
A\left(
\begin{array}{c}
  \lambda_1 \\
  \lambda_2 \\
\end{array}
\right)=0,
\ende
where  $(\lambda_1: \lambda_2)$ are the homogeneous coordinates on $\PP^1$. We simply extend the $\ZZ_p$ action given by (\ref{zetpaction}) to the resolved conifold, clearly this will act trivially on the base $\PP^{1}$ but non-trivially on the fibre co-ordinates since these are contained in the matrix $A$. Thus the resulting orbifold is a particular fibration of an $A_p$-singularity over $\PP^1$. We will show that the large $N$ dual of $\defconip$ is the blowup of this space.

The procedure of blowing up can be convenient described in the language of toric geometry but first we need the fan for the singular manifold. The standard two coordinate patches of $\PP^1$ can be used to trivialize the ${\cal O}_{-1}+{\cal O}_{-1}\rightarrow \projective^1$ bundle. On the first patch we have coordinates $(u, x_1, x_2)=(\lambda_1/\lambda_2,z_1,z_2)$ and on the second we have $(v,y_1, y_2)=(\lambda_2/\lambda_1,-z_3,-z_4)$. On the overlap of the two patches these coordinates are related by $v=1/u, y_1=u x_1, y_2= u x_2$. $\zet_p$ acts on the fiber coordinates. For the quotient of the first patch, therefore, we introduce invariant coordinates $(\xi_1,\xi_2,\xi_3):=(x_1^p, x_2^p, x_1 x_2)$ satisfying $\xi_1\xi_2=\xi_3^p$. Similarly we introduce $(\eta_1,\eta_2,\eta_3)$ satisfying $\eta_1\eta_2=\eta_3^p$ for the second patch. These two sets of coordinates are glued togther by $v=1/u,\eta_1=u^p \xi_1, \eta_2=u^p \xi_2, \eta_3=u^2 \xi_3$. 

Toric geometry is designed to encode these transition functions (see the appendix for notations). Namely, we get the relations 

\be
\begin{array}{ll}
u_{12}+u_{13}=pu_{14},& u_{22}+u_{23}=pu_{24}, \\
u_{21}=-u_{11},& u_{22}=pu_{11}+u_{12},\\ 
u_{23}=p u_{11}+u_{13},& u_{24}=2u_{11}+u_{14}
\end{array}
\ende
for the lattice vectors in $M$. The first two relations imply that $u_{14}$ and $u_{24}$ are not vertices of the dual cones $\check{\sigma}_1$ and $\check{\sigma}_2$ respectively, where $\check{\sigma}_1$ is spanned by $u_{11},u_{12}, u_{13}$, and $\check{\sigma}_2$ is spanned by $u_{21}, u_{22}, u_{23}$. We choose a basis such that $u_{11}=(1,0,0), u_{12}=(0,1,0), u_{14}=(0,0,1)$, this determines all remaining vectors $u_{ij}$.

\begin{figure}[bth!]
\centerline{ \epsfig{file=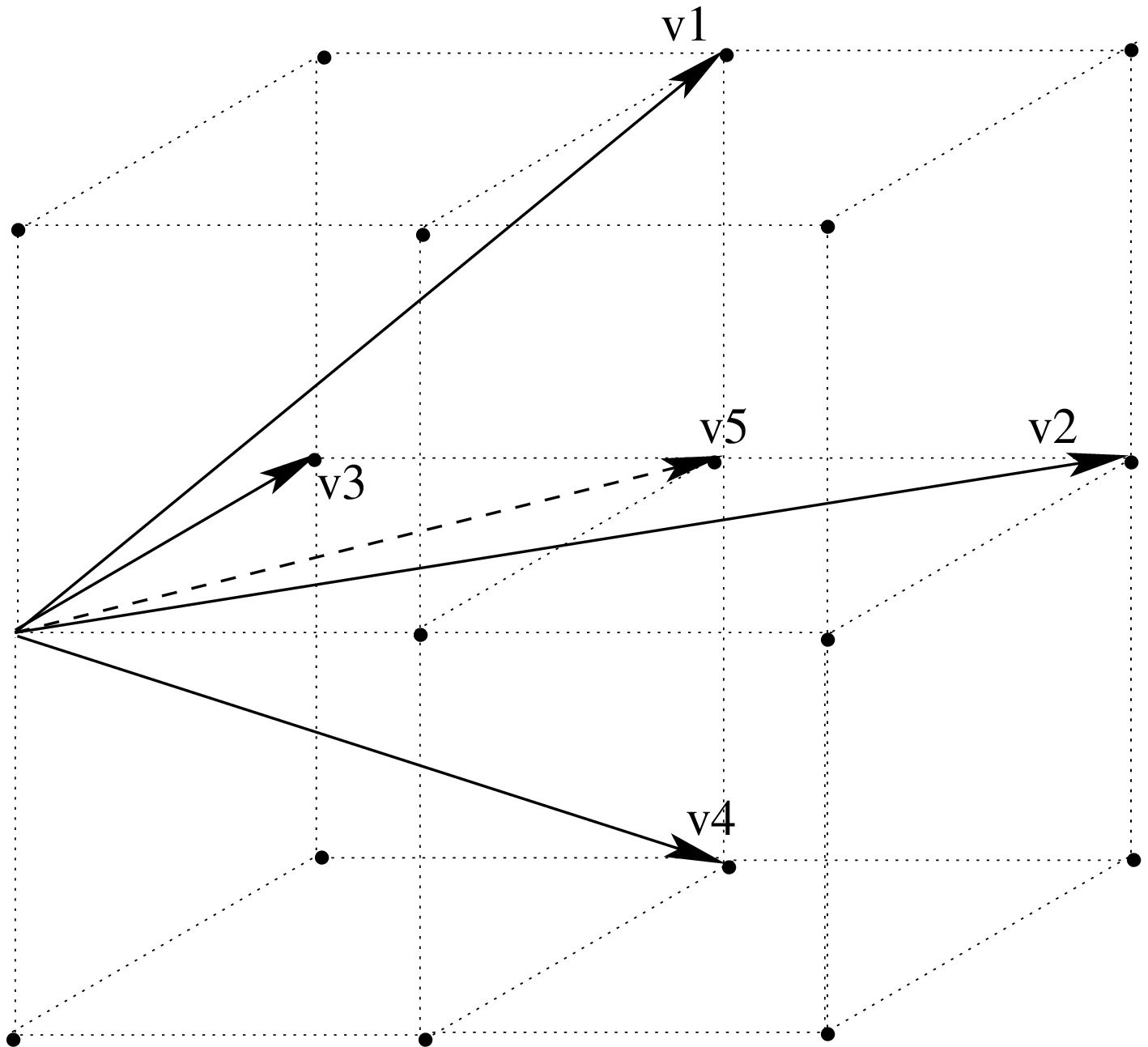,width=9cm,height=9cm}}
\caption{\sl The fan for the resolution of the $\ZZ_{2}$ orbifold of $\resconi$.}
\label{toricp2}
\end{figure}

From this we find that the cone $\sigma_1$ is generated by $v_1, v_2, v_3$, and the cone $\sigma_2$ is generated by $v_4, v_2, v_3$, where $v_1=(1,0,0), v_2=(0,p,1), v_3=(0,0,1), v_4=(-1,p,2)$. After a $GL(3,\ZZ)$ transformation $(x',y',z')=(x+y,x, x+z)$, they become $v_i=(\nu_i,1)$, where $\nu_1=(1,1),\nu_2=(p,0), \nu_3=(0,0), \nu_4=(p-1,-1)$. The toric fan of $A_p\ra \PP^{1}$ consists of the cones $\sigma_1$, $\sigma_2$ and their faces.

\begin{figure}[bth!]
\centerline{ \epsfig{file=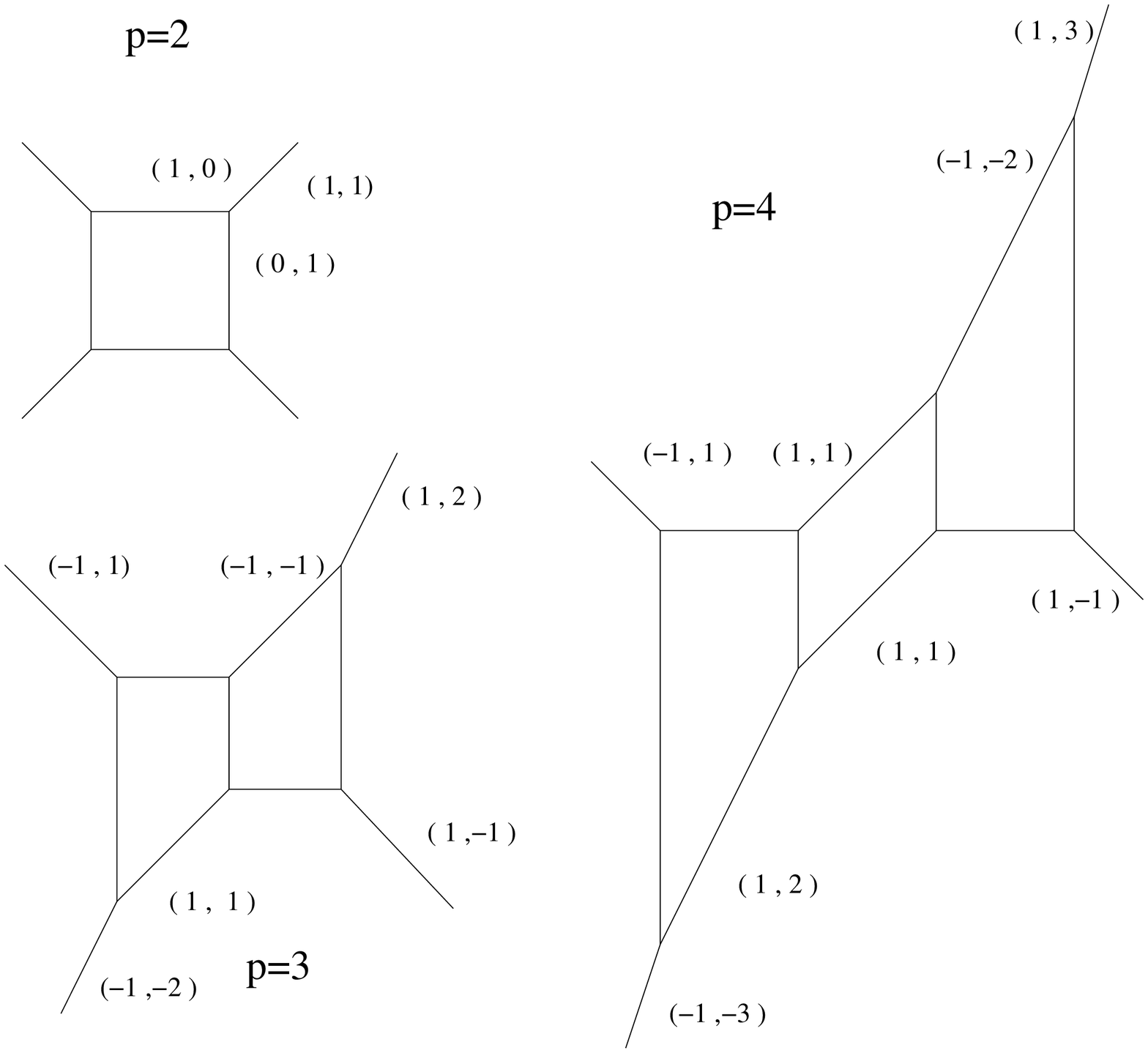,width=10.7cm,height=14cm,angle=-90}}
\caption{\sl The toric web diagrams of the large $N$ dual of $\defconip$ for some values of $p$.}
\label{tord}
\end{figure}

To obtain a smooth manifold, we subdivide the cones $\sigma_1,\sigma_2$ by introducing $v_{i}=(\nu_{i},1)$ where $\nu_{i}=(i-4,0)$ for $i=5,\ldots,p+3$. The fan $\Delta$ of the smooth toric manifold is the union of $[v_1,v_3,v_5]$, $[v_1,v_{p+3},v_2]$, $[v_1,v_5,v_6]$, $[v_1,v_6,v_7]$, $\ldots$, $[v_1,v_{p+2},v_{p+3}]$ from $\sigma_1$ and $[v_4,v_3,v_5]$, $[v_4,v_{p+3},v_2]$, $[v_4,v_5,v_6]$, $[v_4,v_6,v_7]$, $\ldots$, $[v_4,v_{p+2},v_{p+3}]$ from $\sigma_2$ (see figure \ref{toricp2}). Here $[u,v,w]$ denotes the cone spanned by the vectors $u,v,w$. It is easily seen that each cone now has volume equal to $1$, where we normalize the volume of the cone generated by $[1,0,0],[0,1,0],[0,0,1]$ to $1$. The toric web diagrams for these fans are drawn for the cases $p=1,2,3$ in figure \ref{tord}. This geometry and other $A_p$ fibrations over $\PP^{1}$ were recently considered in \cite{Iqbal:2003ix, Iqbal:2003zz} for the purposes of geometric engineering \cite{Katz:1996fh}. Setting $m=0$ in \cite{Iqbal:2003zz} gives the above geometry.

From the data of the fan $\Delta$ we can read off the charge vectors $Q_a, a=0,1,\cdots, p-1$ of the corresponding linear sigma model. They are the generators of the lattice $D=\{Q\in \zet^{p+3}|\sum_i Q_i v_i=0\}$ \cite{Morrison:1994fr}. After rearranging the columns, these are given by

\be
\begin{array}{lrrrrrrrrrrrrrl}
Q_0&=(&0,&1,&1,&-1,&-1,&0,&\ldots,&0,&0,&0,&\cdots &0),\\
Q_j&=(&-(j+1),&j,&0,&0,&0,&0,&\ldots,&0,& \stackrel{(5+j){\rm th}}{1},&0,&\cdots&0),\\
Q_{p-1}&=(&-p,&p-1,&1,&0,&0,&0,&\ldots,&0, &0,&0,&\cdots,&0),
\end{array}
\ende
where $1\leq j\leq p-2$. According to \cite{Hori:2000kt, Aganagic:2000gs}, the mirror geometry is given by $zw=\sum_{i=0}^{p+2}e^{-y_i}$ where the fields are related  by $\sum_iQ_{ai}y_i=t_a$ and one of the $y_i$ variables is to be set to zero. Eliminating $y_2, y_4, y_{4+j}\ (1\leq j\leq p-2)$, and setting $y_1=0$, we get the Riemann surface inside this threefold to be

\be
0=1+e^{-y_0}+e^{-y_3}+e^{-t_{p-1}}e^{-py_0}+e^{t_0-t_{p-1}}e^{-py_0+y_3}
+\sum_{j=1}^{p-2}e^{-t_j}e^{-(j+1)y_0}.
\ende

Now after the co-ordinate transformation $u=-y_0-t_{p-1}/p, v=y_3+t_0+\pi i$, we can write this as

\be \label{matrix_curve}
(e^v-1)(e^{pu+v}-1)+e^{t_0}-1- e^v\left( e^{t_{p-1}/p}e^u+\sum_{j=1}^{p-2}
e^{-t_j+(j+1)t_{p-1}/p} e^{(j+1)u}\right) =0
\ende
which is precisely (\ref{spectral}).


\begin{acknowledgments}
NH would like to thank Jaume Gomis for useful discussions. TO would like to thank Hiroshi Ooguri for collaboration on a related project \cite{Takuya}. We also thank Hiroshi Ooguri for reading the manuscript. This research was supported in part (NH) by the National Science Foundation under
Grant No. PHY99-07949 and (TO) by  DOE grant DE-FG03-92ER40701.
\end{acknowledgments}


\begin{appendix}

\section{Toric variety described by a fan}

A fan $\Delta$ in $N=\zet^n$
is a collection of strongly convex rational polyhedral cones in 
$N_\reals=N\otimes_\zet \reals$,
such that (i) a face of any cone in $\Delta$ is also a cone in $\Delta$
and (ii) the intersection of two cones in $\Delta$ is a face of each cone.

The procedure to construct the charts and transition functions is the following
\cite{Greene:1996cy}.
\begin{enumerate}
\item 
Let $M$ be the dual lattice of $N$. For each cone 
$\sigma_i\in \Delta$ of maximal dimension,
define the dual cone as $\check{\sigma}_i:=\{u\in M_\reals| (u,v)\geq 0, \forall v\in \sigma_i\} $.

\item For each dual cone $\check{\sigma}_i$, choose lattice points $u_{i,j}$ $(j=1,2,\cdots,r_i)$ such that
$\check{\sigma}_i\cap M= \zet_+ u_{i,1}+\cdots+\zet_+ u_{i,r_i}$ ($\zet_+=\{0,1,2,\cdots\})$.

\item For each dual cone $\check{\sigma}_i$, find a set of fundamental relations of 
$u_{i,1}, \cdots,u_{i,r_i}$ in the form
$\sum_{j=1}^{r_i}p_{s,j} u_{i,j}=0$, $s=1,\cdots,R_i$.

\item For each  cone $\sigma_i$, define the patch as $U_{\sigma_i}:=\{(z_{i,1},\cdots,z_{i,r_i})\in
\complex^{r_i}|z_{i,1}^{p_{s,1}}\cdots z_{i,r_i}^{p_{s, r_i}}=1, \forall s\}$.

\item For each pair of cones $\sigma_i, \sigma_j$, find a set of fundamental relations
of $u_{i,1},$ $\ldots,u_{i,r_i},$ $u_{j,1},$ $\ldots,$ $u_{j,r_j}$ in the form
$\sum_{l=1}^{r_i}q_{i,j,l} u_{i,l}$+$\sum_{l=1}^{r_j}q_{i,j,l}' u_{j,l}=0$.

\item Glue the two patches via
$z_{i,1}^{q_{i,j,1}}\cdots z_{i,r_i}^{q_{i,j,r_i}}
z_{j,1}^{q_{i,j,1}'}\cdots z_{j,r_j}^{q_{i,j,r_j } ' }=1$.

\end{enumerate}

\end{appendix}

\end{document}